\documentclass[12pt]{article}
\usepackage{amssymb,amsmath,epsfig}

\begin{document}

\title{\bf Thermodynamics in $f(T)$ Gravity and Corrected Entropies}
\author{M. Sharif \thanks {msharif.math@pu.edu.pk} and
Shamaila Rani\thanks {shamailatoor.math@yahoo.com}\\
Department of Mathematics, University of the Punjab,\\
Quaid-e-Azam Campus, Lahore-54590, Pakistan.}

\date{}

\maketitle
\begin{abstract}
This paper is devoted to study the generalized second law of
thermodynamics in $f(T)$ gravity. We use quantum corrections such as
power-law and logarithmic corrected entropies to the horizon entropy
along with Gibbs' equation in the thermal equilibrium state. We
derive $f(T)$ model by taking into account a power-law scale factor
through the first modified Friedmann equation which obeys the
condition for a realistic model. Two types of horizons, i.e., Hubble
and event horizons are used to check the validity of the generalized
second law of thermodynamics with corrected entropies. We conclude
that this law holds with a specific range of entropy parameter on
both horizons in the case of power-law corrected entropy, while it
violates for all values of entropy parameter on both horizons for
logarithmic corrected entropy.
\end{abstract}
\textbf{Keywords:} $f(T)$ gravity; Generalized second law of thermodynamics.\\
\textbf{PACS:} 04.50.kd; 05.70.-a.

\section{Introduction}

The modified theories of gravity are well-known for a combined
motivation coming from cosmology, high-energy physics, astrophysics
and quantum physics. In these theories, the $f(T)$ gravity
\cite{1}-\cite{1b} has attained a lot of interest for describing the
present accelerated state of the universe along with some attractive
features geometrically and physically. It is a straightforward
modification of teleparallel theory \cite{2}-\cite{2b} by
introducing an arbitrary function $f(T)$ in its Lagrangian density
instead of torsion scalar $T$. This theory deals with torsion via
Weitzenb\"{o}ck connection (having zero curvature) instead of
Levi-Civita connection responsible for curvature. It is easy to
tackle in the sense that the torsion scalar involves products of
first derivatives of tetrad (dynamical field) and the field
equations contain the second order.

This modified gravity has been studied extensively under many
phenomena, such as, accelerated expansion of the universe
\cite{3}-\cite{3c}, large-scale structure \cite{4}, cosmological
perturbations \cite{5,5a}, discussion of Birkhoff's theorem
\cite{6}, static spherically symmetric solutions \cite{7}, solar
system constraints \cite{8}, reconstructions via scalar fields
\cite{9,9a}, viability of models through cosmographic technique
\cite{10,10a}, thermodynamics \cite{11}-\cite{11++} and many more.
The connection between thermodynamics and gravitation is served by
the black hole thermodynamics \cite{12,12a}. Jacobson \cite{13} used
the relation among the entropy and horizon area in thermodynamics
and derived the Einstein equations. This work is extended for the
curvature correction \cite{g} to the entropy in the form of
polynomial Ricci scalar in non-equilibrium thermodynamics. Bamba et
al. \cite{h} investigated that the equations of motion of modified
gravity theories, particularly $f(R),~f(G)$, scalar Gauss-Bonnet and
non-local theories are equivalent to to the Clausius relation in
thermodynamics. It is worthwhile to check the viability of the
generalized second law of thermodynamics (GSLT) in the accelerated
universe \cite{14}-\cite{14c}.

Bamba and Geng \cite{11} explored the thermodynamics in equilibrium
and non-equilibrium descriptions for apparent horizon in $f(T)$
gravity. Karami and Abdolmaleki \cite{11+} explored the validity of
GSLT on Hubble horizon using power-law and exponential models. They
concluded that this law holds for both models from early to present
universe, while it is violated in the future epoch. Bamba et al.
\cite{15} discussed the finite time singularities, Little Rip,
Pseudo-Rip cosmologies and thermodynamics for the apparent horizon
bounded universe. Some people have discussed the validity of this
law by introducing correction terms in entropy and horizon area in
general relativity as well as in modified gravities.

Debnath et al. \cite{16} investigated the validity of GSLT by taking
power-law corrected entropy (PLCE) in equilibrium and
non-equilibrium cases for apparent and event horizons in general
relativity. They found some constraints on the power-law parameter
and small perturbation in de Sitter spacetime for the validity of
this law. In the case of logarithmic corrected entropy (LCE),
Sadjadi and Jamil \cite{17} found that the validity occurs for
positive LCE parameter and concluded that this law holds throughout
the universe for spatial curvature with any dark energy model.
Sharif and Jawad \cite{18} discussed the validity of GSLT with
corrected entropies for three different systems in the closed
universe. Recently, Bamba et al. \cite{11++} studied the constraints
on PLCE and LCE parameters to satisfy or violate the GSLT in $f(T)$
gravity by taking a system of n-component fluids in thermal
equilibrium for apparent and event horizons.

In this paper, we discuss the validity of this law with PLEC and LEC
for Hubble as well as event horizons in $f(T)$ gravity by
constructing a realistic $f(T)$ model. The scheme of the paper is as
follows. In section \textbf{2}, we provide a brief review of $f(T)$
formalism and entropy corrections. Section \textbf{3} is devoted to
discuss the validity of GSLT with PLCE and LCE for Hubble as well as
event horizons in equilibrium state. The last section summarizes the
results.

\section{Brief Review}

Here we provide briefly the formalism of $f(T)$ gravity, its field
equations and some entropy corrections to the horizon entropy.

\subsection{$f(T)$ Gravity and Field Equations}

The two connected parts/structures of a manifold involve Riemannian
structure with a definite metric and non-Riemannian structure having
torsion or non-metricity. The Weitzenb\"{o}ck spacetime is defined
by the second structure which has zero Riemannian tensor but
non-zero torsion based on the tetrad field. This was originally
proposed by Einstein to unify the electromagnetism with gravity and
introduced teleparallel theory of gravity. The dynamical tetrad
field $h_a(x^{\mu})$ is an orthonormal basis for the tangent space
at each point of the manifold \cite{f}. This field is analyzed by
tetrad components $h^{\mu}_{a}$ ($\mu,a=0,1,2,3$) in the coordinate
basis $h_a=h^{\mu}_{a}\partial_{\mu}$, related by
$h_{\mu}^{a}h^{\mu}_{b}=\delta^{a}_{b}$ and
$h_{\mu}^{a}h^{\nu}_{a}=\delta^{\nu}_{\mu}$. We denote the
coordinates on the manifold by Greek indices while the Latin
alphabets refer to the tangent space. The metric tensor is obtained
by the dual tetrad components as
$g_{\mu\nu}=\eta_{ab}h_{\mu}^{a}h_{\nu}^{b}$.

The Weitzenb\"{o}ck connection is defined from tetrad as
${\tilde{\Gamma}^\rho}_{~\mu\nu}=h^{\rho}_{a}\partial_{\nu}h^{a}_{\mu}$,
yielding the following antisymmetric torsion tensor
\begin{equation}\nonumber
{T^\rho}_{\mu\nu}={\tilde{\Gamma}^\rho}_{~\nu\mu}-{\tilde{\Gamma}^\rho}_{~\mu\nu}=h^{\rho}_{a}
(\partial_{\mu}h^{a}_{\nu}-\partial_{\nu}h^{a}_{\mu}).
\end{equation}
The antisymmetric superpotential ${S_\rho}^{\mu\nu}$ and contorsion
${K^{\mu\nu}}_{\rho}$ tensors are
\begin{eqnarray}\nonumber
{S_\rho}^{\mu\nu}=\frac{1}{2}({K^{\mu\nu}}_{\rho}
+\delta^{\mu}_{\rho}{T^{\theta\nu}}_{\theta}-\delta^{\nu}_{\rho}{T^{\theta\mu}}_{\theta}),\quad
{K^{\mu\nu}}_{\rho}&=&-\frac{1}{2}({T^{\mu\nu}}_{\rho}
-{T^{\nu\mu}}_{\rho}-{T_\rho}^{\mu\nu}),
\end{eqnarray}
which are used to define the torsion scalar as
\begin{equation}\label{1}
T={T^\rho}_{\mu\nu}{S_\rho}^{\mu\nu}.
\end{equation}
The action of $f(T)$ gravity is given by
\begin{equation}\label{2}
S=\frac{1}{2\kappa^2}\int d^{4}x[hf(T)+\mathcal{\mathcal{L}}_m],
\end{equation}
where $h=\sqrt{-g},~\kappa^{2}=8\pi G,~G$ is the gravitational
constant and $\mathcal{L}_m$ is the matter Lagrangian density inside
the universe. The corresponding field equations are obtained by
varying this action with respect to tetrad as \cite{20}
\begin{equation}\label{3}
[h^{-1}\partial_{\mu}(hS_{a}~^{\mu\nu})
+h^{\lambda}_{a}T^{\rho}~_{\mu\lambda}S_{\rho}~^{\nu\mu}]f_{T}
+S_{a}~^{\mu\nu}\partial_{\mu}(T)
f_{TT}+\frac{1}{4}h^{\nu}_{a}f=\frac{1}{2}\kappa^{2}h^{\rho}_{a}T^{\nu}_{\rho},
\end{equation}
where $f_T=df/dT,~f_{TT}=d^{2}f/dT^{2}$ and $T^{\nu}_{\rho}$ is the
energy-momentum tensor of perfect fluid.

For the flat FRW universe, we take tetrad components as
$h^{a}_{\nu}=diag(1,a,a,a)$ \cite{9,9a}. The corresponding modified
Friedmann equations are
\begin{eqnarray}\label{6}
12H^2f_T+f&=&2\kappa^{2}\rho,\\\label{6+}
48\dot{H}H^2f_{TT}-(12H^2+4\dot{H})f_T-f&=&2\kappa^{2}p,
\end{eqnarray}
where $\rho$ and $p$ are the total energy density and pressure of
the universe and $H=\frac{\dot{a}}{a}$. The above field equations
can be written as
\begin{eqnarray}\label{7}
\frac{3H^2}{\kappa^2}=\rho,\quad -\frac{2\dot{H}}{\kappa^2}=\rho+p,
\end{eqnarray}
where $\rho=\rho_m+\rho_T$ and $p=p_m+p_T$. We assume here the
pressureless (dust) matter, i.e., $p_m=0$ and $\rho_T,~p_T$ are
torsion contributions given by
\begin{eqnarray}\label{8}
\rho_T&=&\frac{1}{2\kappa^2}(-12H^2f_T-f+6H^2),\\\label{9}
p_T&=&-\frac{1}{2\kappa^2}(48\dot{H}H^2f_{TT}-(12H^2+4\dot{H})f_T-f+6H^2+4\dot{H}).
\end{eqnarray}
The corresponding energy conservations are
\begin{equation}\label{10}
\dot{\rho}_m+3H\rho_m=0,\quad \dot{\rho}_T+3H(\rho_T+p_T)=0.
\end{equation}
For dust matter, it yields
\begin{equation}\label{11}
\rho_{m}=\rho_{m0}a^{-3},
\end{equation}
where $\rho_{m0}$ is an arbitrary constant.

We assume here the power-law scale factor as \cite{21}-\cite{21b}
\begin{equation}\label{12}
a(t)=a_{0}(t_{s}-t)^{-b},\quad b>0, \quad t_{s}\geq t,
\end{equation}
where $a_0$ is the present-day value of the scale factor. This scale
factor indicates the superaccelerated universe with a Big Rip
singularity at $t=t_{s}$. Using this scale factor, the Hubble
parameter, torsion scalar and $\dot{H}$ become
\begin{equation}\label{13}
H=\frac{b}{t_s-t},\quad T=-\frac{6b^2}{(t_s-t)^{2}},\quad
\dot{H}=\frac{b}{(t_s-t)^{2}}.
\end{equation}
Inserting these values in Eq.(\ref{6}), we obtain
\begin{eqnarray}\label{14}
f(T)=c\left(-\frac{T}{6b^2}\right)^{\frac{1}{2}}+\frac{2\kappa^2\rho_{m0}}{a_{0}^3(3b+1)}
\left(-\frac{6b^2}{T}\right)^{\frac{3b}{2}},
\end{eqnarray}
where $c$ is an integration constant which can be determined by
imposing a suitable boundary condition. This model satisfies the
condition of a realistic model, i.e., $\frac{f}{T}\rightarrow 0$
\cite{10,10a} as $T\rightarrow\infty$ at high redshift, representing
an accelerated expansion of the universe which is consistent with
the primordial nucleosynthesis and cosmic microwave background
constraints. To determine $c$, we impose the condition on
gravitational constant $G$. For non-linear $f(T)$, Eq.(\ref{6})
implies an effective gravitational constant (time dependent),
$G_{eff}$ instead of $G$ ($\kappa^2=8\pi G$). It must reduce to the
present day value of $G$ for linear $f(T)$ which yields the
condition $f_T(T_0)=1$, where $T_0=-6H_0^2$ and $H_0$ is the present
day value of Hubble parameter. Using the model (\ref{14}) in this
condition, it follows that
\begin{equation}\label{15}
c=12bH_0\left[\frac{b\kappa^2\rho_{m0}}{2a_{0}^3(3b+1)H_0^2}
\left(\frac{b}{H_0}\right)^{3b}-1\right].
\end{equation}

\subsection{Corrected Entropies}

The correction terms in the entropy-area relationship are widely
discussed to study the thermodynamical systems. The entropy of the
horizon is proportional to the area of the horizon ($S\propto A$) in
the Einstein gravity. If we modify the action of gravity theory by
adding some extra curvature terms, it changes the entropy-area
relation, e.g., it takes the form $S\propto Af_R$ \cite{c} in $f(R)$
theory, where $f_R$ is the derivative of arbitrary function $f$ with
respect to the Ricci scalar $R$. This relationship is affected by
some field anomalies and gravitational fluctuations based on black
hole physics. To deal with these fluctuations, quantum corrections
to the semi-classical entropy law have been introduced in the form
of power-law and logarithmic.

The entanglement of quantum fields in and out the horizon generate
the corrections to the entropy such as a power-corrected area term
in the entropy expression. The power-law corrected entropy takes the
form \cite{11++,22}
\begin{equation}\label{23}
S_X=\frac{A}{4G}\left(1-K_{\alpha}A^{1-\frac{\alpha}{2}}\right),\quad
K_{\alpha}=\frac{\alpha(4\pi)^{\frac{\alpha}{2}-1}}{(4-\alpha)r_c^{2-\alpha}}
\end{equation}
where $A=4\pi R_{X}^2,~R_X$ is the radius of an arbitrary horizon
$X$, $\alpha$ is a dimensionless constant which should be greater
than zero for entropy to be well-defined and $r_c\sim H_0^{-1}$ is
the crossover scale \cite{a}. The correction term in Eq.(\ref{23})
is the result of wave function of the field which is the
entanglement of ground and excited states. The excited state
contributes to the correction while the ground state entanglement
entropy satisfies the black hole entropy-area relationship. For
higher excitation states, the correction term is more significant
and it falls off rapidly with increments in area, i.e., in the
semi-classical limit (large area), the entropy-area law is
recovered. The curvature correction in the Einstein-Hilbert action
is formed due to quantum corrections into the entropy-area
relationship and vice versa. This leads to the logarithmic corrected
entropy \cite{11++,23,23a}
\begin{equation}\label{24}
S_X=\frac{A}{4G}+\beta \log\left(\frac{A}{4G}\right)+\gamma,
\end{equation}
where $\beta$ and $\gamma$ are dimensionless constants whose exact
values are not yet known. These corrections arise due to
mass-charge, quantum and thermal equilibrium fluctuations in loop
quantum gravity.

\section{Thermodynamics}

The generalized second law of thermodynamics states that the sum of
entropy of the horizon and entropy of total matter inside the
horizon does not decrease with time. The Clausius relation using the
first law of thermodynamics is found to be $-dE=T_{X}dS_X$, where
$S_{X}=\frac{A}{4G}$ is the Bekenstein entropy (entropy-area
relation) and $T_{X}=\frac{1}{2\pi R_{X}}$ is the Hawking
temperature. Miao et al. \cite{b} found that the first law of
thermodynamics violates in $f(T)$ gravity due to lack of local
Lorentz invariance which leads to some degrees of freedom and
results an additional entropy production term $S_{P}$. In order to
reduce the degrees of freedom, they get a condition
$\epsilon=\frac{4f_{TT}(0)}{f(0)}<0$, where $\epsilon$ denotes the
violation of local Lorentz invariance. This parameter $\epsilon$ and
$f_{TT}$ should be very small to be consistent with the experiments.
Thus, the first law of thermodynamics holds if $f_{TT}$ is very
small and entropy horizon becomes $S_{X}=\frac{Af_T}{4G}$ with zero
$S_P$. The entropy-area relation is modified to $A\rightarrow Af_T$,
which leads to the modification of power-law (PX) and logarithmic
(LX) corrected entropies (\ref{23}) and (\ref{24}), given by
\cite{11++}
\begin{equation}\label{16}
S_{PX}=\frac{Af_T}{4G}\left(1-K_{\alpha}(Af_T)^{1-\frac{\alpha}{2}}\right),\quad
S_{LX}=\frac{Af_T}{4G}+\beta \log
\left(\frac{Af_T}{4G}\right)+\gamma.
\end{equation}

The Gibbs' equation is used to find the rate of change of normal
entropy $S_{I}$ of the horizon
\begin{equation}\label{17}
\dot{S}_I+\dot{S}_{P}=\frac{1}{T_X}\left(\frac{dE_{I}}{dt}+p\frac{dV}{dt}\right),
\end{equation}
where $E_{I}=\rho V,~V=\frac{4}{3}\pi R_{X}^3$ is the volume of the
horizon. Inserting these values in this equation, it follows that
\begin{equation}\label{18}
\dot{S}_I+\dot{S}_{P}=8\pi^2 R_X^3(\dot{R}_X-HR_X)(\rho+p).
\end{equation}
The time derivative of power-law and logarithmic corrected entropies
become
\begin{eqnarray}\label{19}
\dot{S}_{PX}&=&\frac{2\pi
R_X}{G}(\dot{R}_Xf_T-6H\dot{H}R_Xf_{TT})[1-(2-\frac{\alpha}{2})K_{\alpha}(4\pi
R_X^2 f_T)^{1-\frac{\alpha}{2}}], \\\label{20}
\dot{S}_{LX}&=&2\left(\frac{\pi R_X}{G}+\frac{\beta}{R_X
f_T}\right)(\dot{R}_Xf_T-6H\dot{H}R_Xf_{TT}).
\end{eqnarray}
For the validity of GSLT, we first see the behavior of the second
derivative of the model (\ref{14}) with $a(t)=\frac{a_0}{1+z}$, it
follows that
\begin{eqnarray}\label{21}
f_{TT}=\frac{\kappa^2\rho_{m0}(3b+2)}{24a_{0}^3b^3(3b+1)}(1+z)^{3+\frac{4}{b}}-
\frac{c}{144b^4}(1+z)^{\frac{3}{b}}.
\end{eqnarray}
Its plot versus $z$ is shown in Figure \textbf{1} for $b=2,3,5$ and
using $\kappa^2=\rho_{m0}=a_0=1,~H_0=74.2 km s^{-1} Mpc^{-1}$. The
graph indicates that $f_{TT}\ll1$ in the range $z<0.18$ for $b=2$
whereas it satisfies for $b=3,5$ and all values of $z$. Thus we take
the entropy production term to be zero in Eq.(\ref{18}).
\begin{figure} \centering
\epsfig{file=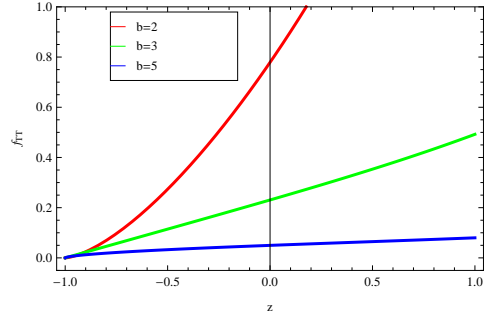,width=.50\linewidth}\caption{Plot of $f_{TT}$
versus $z$.}
\end{figure}

In the following, we check the validity of GSLT for Hubble and event
horizons for both corrected entropies.

\subsection{Hubble Horizon}

Consider the boundary of thermal system of the FRW spacetime covered
by the Hubble horizon in equilibrium state. It is the reduction of
apparent horizon for flat space \cite{24}. The radius of Hubble
horizon and its time derivative are given by
\begin{equation}\label{22}
R_H=\frac{1}{H},\quad \dot{R}_H=-\frac{\dot{H}}{H^2}.
\end{equation}

\subsubsection*{Power-law Corrected Entropy}

Replacing $X$ by $H$ in Eqs.(\ref{18}) and (\ref{19}), the time
derivative of total entropy for Hubble horizon, i.e.,
$\dot{S}_{PLCE}=\dot{S}_{PH}+\dot{S}_{I}$ becomes
\begin{eqnarray}\nonumber
\dot{S}_{PLCE}&=&-\frac{8\pi^2}{H^3}\left(1+\frac{\dot{H}}{H^2}\right)\left[\rho_{m0}a^{-3}+\frac{1}{4\pi
G}(2\dot{H}Tf_{TT}+\dot{H}(f_T-1))\right]\\\label{35}&-&\frac{2\pi}{G
H}\left(\frac{\dot{H}}{H^2}f_T+6\dot{H}f_{TT}\right)\left[1-(2-\frac{\alpha}{2})K_{\alpha}\left(\frac{4\pi}{H^2}
f_T\right)^{1-\frac{\alpha}{2}}\right].
\end{eqnarray}
Using Eqs.(\ref{12})-(\ref{14}), we obtain
\begin{eqnarray}\nonumber
\dot{S}_{PLCE}&=&-\frac{\pi (1+z)^{\frac{1}{b}}}{Gb^3}
\left[2(1+b)-\frac{\kappa^2
\rho_{m0}(4+3b)(1+z)^{3+\frac{2}{b}}}{2a_0^3(1+3b)}+\frac{c(1+z)^{\frac{1}{b}}}{4b}\right.
\\\label{36}&\times&\left.\left(1-\frac{\alpha (4\alpha)^{\frac{\alpha}{2}-1}}{2H_0^{\alpha-2}}
(\frac{2\pi \kappa^2
\rho_{m0}(1+z)^{3+\frac{4}{b}}}{a_0^3b^3(1+3b)}-\frac{c\pi
(1+z)^{\frac{3}{b}}}{3b^4})^{1-\frac{\alpha}{2}}\right)\right],
\end{eqnarray}
which is the time derivative of the total entropy with power-law
correction for Hubble horizon with $f(T)$ model (\ref{14}) in terms
of $z$. Its plot versus $z$ and $\alpha$ is shown in Figure
\textbf{2} (left) keeping the same values of constants with $b=3$.
Initially, the graph represents large positive values of
$\dot{S}_{PLCE}$ for higher values of $z$, then it decays and
remains positive for the present universe ($z=0$) towards future
($z<0$). For $\alpha\leq2$, the graph remains positive, otherwise
shows negative behavior. Thus, the GSLT is valid for all values of
$z$ with $\alpha\leq2$, while it violates for $\alpha>2$.
\begin{figure} \centering
\epsfig{file=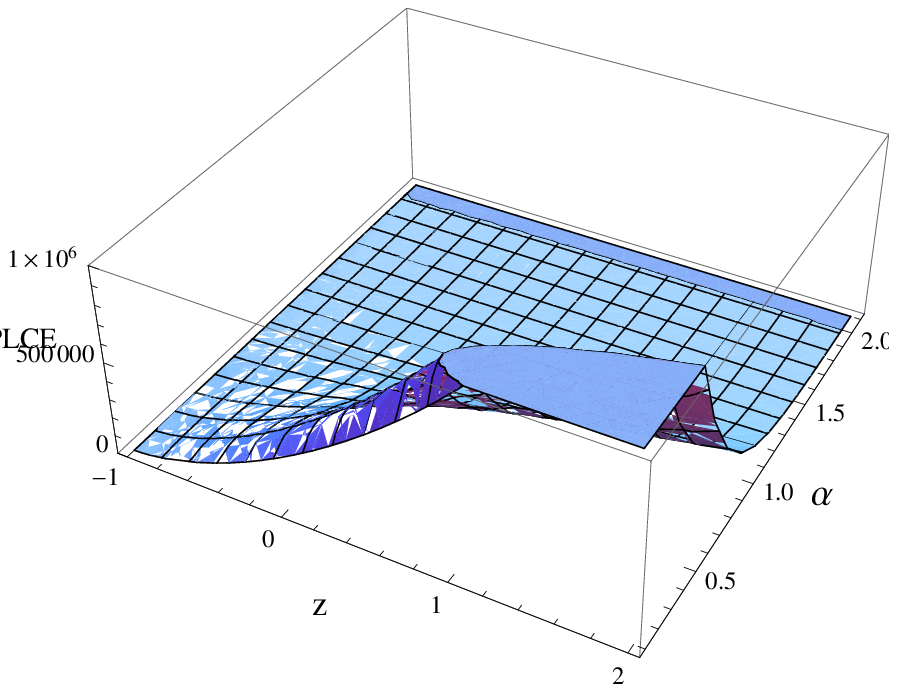,width=.50\linewidth}\epsfig{file=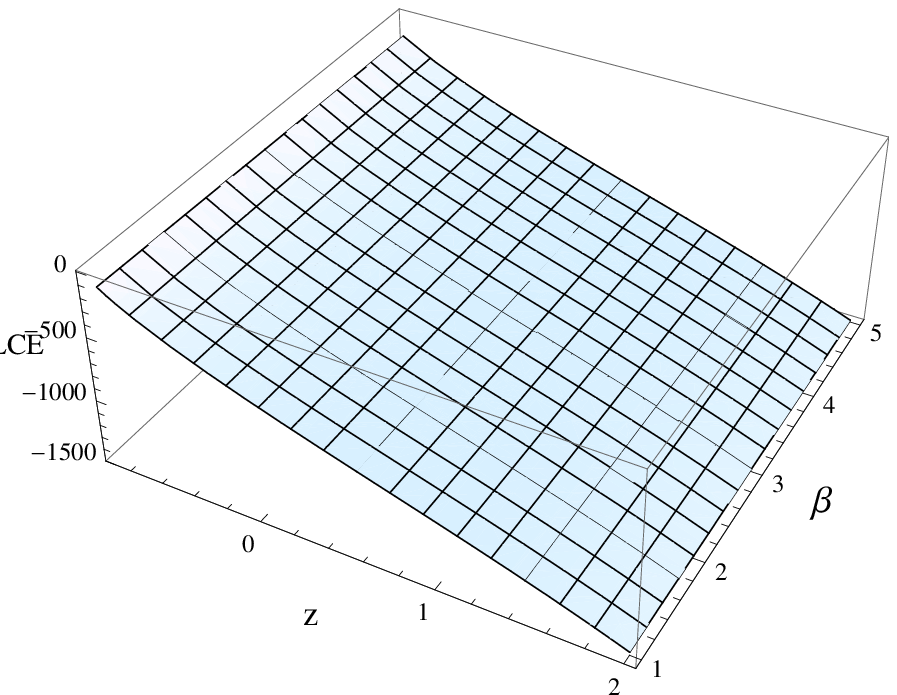,width=.50\linewidth}\caption{Plots
of the rate of change of total entropy versus redshift and model
parameters for Hubble horizon. The left graph is for power-law
corrected entropy and the right is for logarithmic corrected
entropy.}
\end{figure}

\subsubsection*{Logarithmic Corrected Entropy}

The time derivative of total entropy for Hubble horizon
($X\rightarrow H$) with logarithmic correction using Eqs.(\ref{18})
and (\ref{20}), i.e., $\dot{S}_{LCE}=\dot{S}_{LH}+\dot{S}_{I}$ takes
the form
\begin{eqnarray}\nonumber
\dot{S}_{LCE}&=&-2\left(\frac{\pi}{GH}+\frac{\beta
H}{f_T}\right)\left(\frac{\dot{H}}{H^2}f_T-6\dot{H}f_{TT}\right)-\frac{8\pi^2}{H^3}
\left(1+\frac{\dot{H}}{H^2}\right)\\\label{37}&\times&
\left[\rho_{m0}a^{-3}+\frac{1}{4\pi
G}(2\dot{H}Tf_{TT}+\dot{H}(f_T-1))\right].
\end{eqnarray}
Inserting $f(T)$ model along with scale factor and Hubble horizon in
the above equation, we obtain
\begin{eqnarray}\nonumber
\dot{S}_{LCE}&=&-2\left[\frac{\pi (1+z)^{\frac{1}{b}}}{G
b}+\frac{\beta b^2}{(1+z)^{\frac{2}{b}}}\left(\frac{\kappa^2
\rho_{m0}}{2a_{0}^{3}(1+3b)}(1+z)^{3+\frac{1}{b}}-\frac{c}{12b}\right)^{-1}\right]\\\label{37}
&\times&\left[\frac{\kappa^2
\rho_{m0}(4+3b)}{4a_{0}^{3}b^2(1+3b)}(1+z)^{3+\frac{2}{b}}-\frac{c(1+z)^{\frac{1}{b}}}{8b^3}\right]
+\frac{2\pi (1+b)(1+z)^{\frac{1}{b}}}{Gb^3}.
\end{eqnarray}
Figure \textbf{2} (right) represents its behavior versus $z$ and
$\beta$. It indicates negative behavior for all values of $z$ and
$\beta$. As $z$ decreases, $\dot{S}_{LEC}$ increases and gets closer
to positive values for present and future epochs but remains
negative. Thus, GSLT does not hold for logarithmic entropy
correction.

\subsection{Event Horizon}

Now we assume the event horizon \cite{25} as boundary of thermal
equilibrium system whose existence is related to the convergence of
the following integral
\begin{equation}\label{38}
R_{E}=a\int^{\infty}_{t} \frac{dt}{a},\quad \dot{R}_{E}=HR_{E}-1.
\end{equation}
It is the distance of light traveling from present time to infinity
and we replace $\infty$ by $t_s$ for Big Rip future time
singularity.

\subsubsection*{Power-law Corrected Entropy}

Using Eqs.(\ref{18}) and (\ref{19}), replacing the arbitrary horizon
$X$ by event horizon and adding the resulting equations, it yields
\begin{eqnarray}\nonumber
\dot{S}_{PLCE}&=&-8\pi^2(a\int^{\infty}_{t}
\frac{dt}{a})^3\left(\rho_{m0}a^{-3}+\frac{1}{4\pi
G}(2\dot{H}Tf_{TT}+\dot{H}f_T-\dot{H})\right)
\end{eqnarray}
\begin{eqnarray}\nonumber
&+&\frac{2\pi}{G}a\int^{\infty}_{t}
\frac{dt}{a}\left[(Ha\int^{\infty}_{t}
\frac{dt}{a}-1)f_T-6a\int^{\infty}_{t}
\frac{dt}{a}H\dot{H}f_{TT}\right]\\\label{39}&\times&\left[1-(2-\frac{\alpha}{2})K_{\alpha}\left(4\pi
(a\int^{\infty}_{t} \frac{dt}{a})^2
f_T\right)^{1-\frac{\alpha}{2}}\right].
\end{eqnarray}
This is the time derivative of total entropy with power-law
corrected entropy for event horizon. Using $f(T)$ model, $a(t)$ and
$H$ in terms of $z$, we obtain
\begin{eqnarray}\nonumber
\dot{S}_{PLCE}&=&\frac{2\pi
b}{G(1+b)^3}(1+z)^{\frac{1}{b}}-\left[\frac{\pi\kappa^2\rho_{m0}(4+3b)}{2Ga_{0}^{3}b(1+b)^2(1+3b)}
(1+z)^{3+\frac{3}{b}}\right.\\\nonumber&-&\left.\frac{2\pi
c(1+z)^{\frac{2}{b}}}{8Gb^2(1+b)^2}\right]\left[1-\frac{\alpha
(4\alpha)^{\frac{\alpha}{2}-1}}{2H_0^{\alpha-2}}\left(\frac{2\pi
\kappa^2
\rho_{m0}(1+z)^{3+\frac{4}{b}}}{a_0^3b(1+b)^2(1+3b)}\right.\right.\\\label{40}&-&\left.\left.\frac{c\pi
(1+z)^{\frac{3}{b}}}{3b^2(1+b)^2}\right)^{1-\frac{\alpha}{2}}\right].
\end{eqnarray}
Its plot versus $z$ and $\alpha$ is shown in the left panel of
Figure \textbf{3} which expresses the same behavior as for the
Hubble horizon. The only difference is the values of time derivative
of total entropies in the corresponding intervals of $z$. The GSLT
satisfies for all values of $z$ with $\alpha\leq2$ for power-law
corrected entropy.

\subsubsection*{Logarithmic Corrected Entropy}
\begin{figure} \centering
\epsfig{file=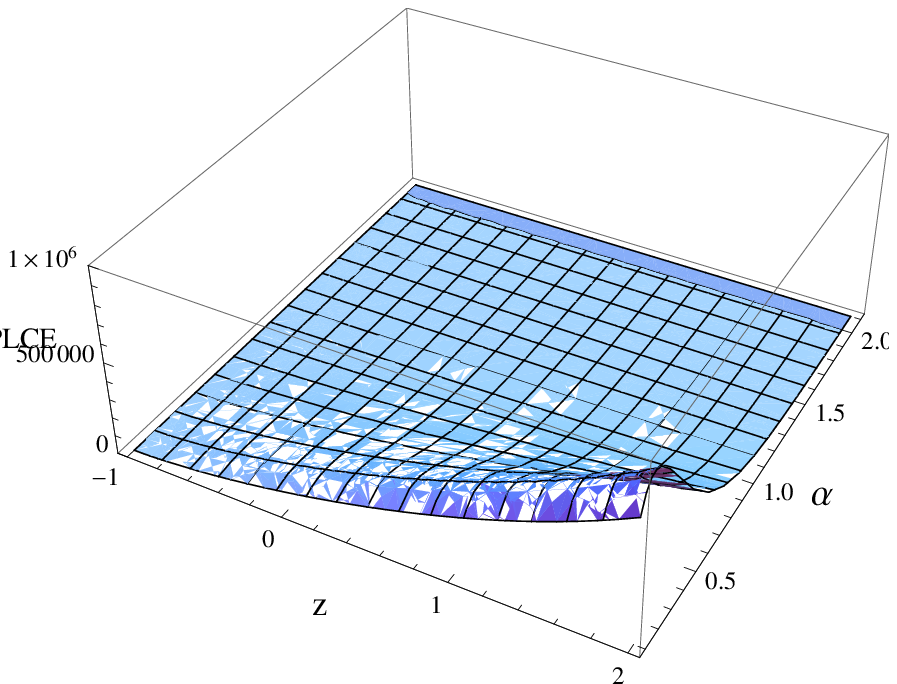,width=.50\linewidth}\epsfig{file=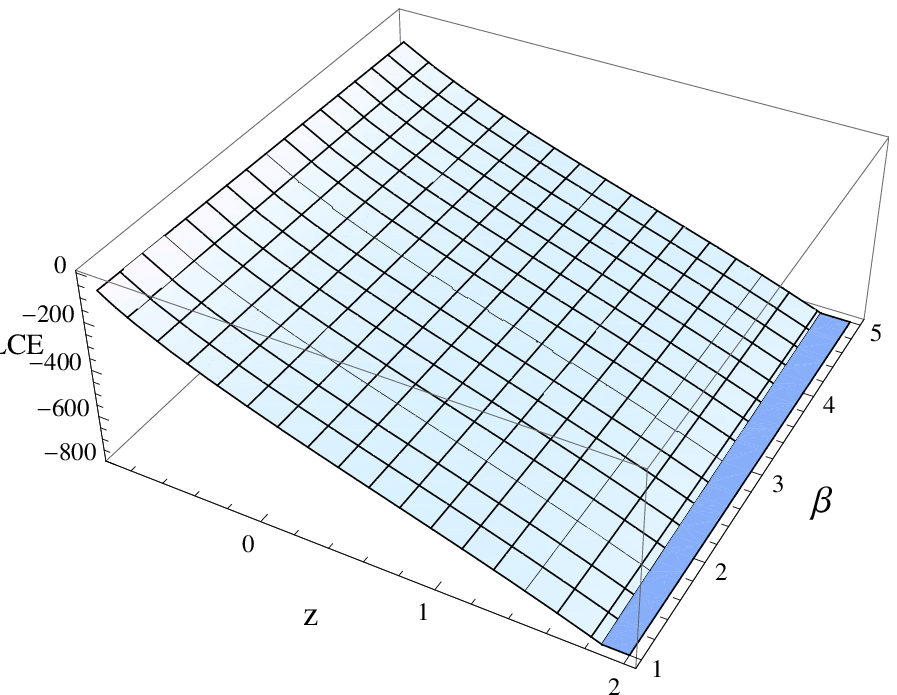,width=.50\linewidth}\caption{Plots
of the rate of change of total entropy versus redshift and model
parameters for event horizon. The left graph is for power-law
corrected entropy and the right is for logarithmic corrected
entropy.}
\end{figure}

For event horizon ($X\rightarrow E$ in Eqs.(\ref{18}) and
(\ref{20})), the rate of change of total entropy becomes
\begin{eqnarray}\nonumber
\dot{S}_{PLCE}&=&-8\pi^2(a\int^{\infty}_{t}
\frac{dt}{a})^3\left(\rho_{m0}a^{-3}+\frac{1}{4\pi
G}(2\dot{H}Tf_{TT}+\dot{H}f_T-\dot{H})\right)\\\nonumber&+&2\left(\frac{\pi
a}{G}\int^{\infty}_{t} \frac{dt}{a}+\frac{\beta}{a\int^{\infty}_{t}
\frac{dt}{a}f_T}\right)\left[(Ha\int^{\infty}_{t}
\frac{dt}{a}-1)f_T-6aH\right.\\\label{41}&\times&\left.
\int^{\infty}_{t} \frac{dt}{a}\dot{H}f_{TT}\right],
\end{eqnarray}
yielding
\begin{eqnarray}\nonumber
\dot{S}_{LCE}&=&-2\left[\frac{\pi (1+z)^{\frac{1}{b}}}{G
(1+b)}+\frac{2b(1+b)\beta}{(1+z)^{\frac{2}{b}}}\left(\frac{\kappa^2
\rho_{m0}}{a_{0}^{3}(1+3b)}(1+z)^{3+\frac{1}{b}}-\frac{c}{6b}\right)^{-1}\right]\\\label{42}&\times&
\left[\frac{\kappa^2
\rho_{m0}(4+3b)(1+z)^{3+\frac{2}{b}}}{4a_{0}^{3}b(1+b)(1+3b)}-\frac{c(1+z)^{\frac{1}{b}}}{8b^2(1+b)}\right]+
\frac{2\pi b(1+z)^{\frac{1}{b}}}{G(1+b)^3}.
\end{eqnarray}
Figure \textbf{3} (right) shows its graph versus $z$ and $\beta$.
This also represents the same behavior of total entropy for the
logarithmic correction with the same range of $z$ and $\beta$ as for
the Hubble horizon. Thus GSLT also violates for the total entropy
having logarithmic correction for the event horizon.

\section{Concluding Remarks}

In this paper, we have discussed the validity of GSLT in the context
of $f(T)$ gravity in FRW universe. We have taken the corrected
entropies such as, PLCE and LCE to the entropy-area relationship. A
power-law scale factor is chosen to construct the $f(T)$ model and
integration constant is found through a boundary condition on
$G_{eff}$. This model satisfies the condition for a realistic model.
We have checked the behavior of the second derivative of $f(T)$
model in order to meet the first law of thermodynamics. The validity
of GSLT with corrected entropies are investigated through graphical
representation for two horizons, Hubble and event horizons in
equilibrium state. The results for both these horizons are
summarized as follows.

The $f(T)$ model satisfies the condition, $f_{TT}\ll1$ which leads
to take entropy production term to be zero. The time derivative of
total entropy with PLCE for Hubble and event horizons represents
positive behavior versus $z$ for $h=3$ within a specific range of
PLCE parameter $\alpha\leq2$. The only difference comes in the
values of PLCE in the corresponding intervals of $z$. Thus GSLT for
this corrected entropy satisfies in the underlying scenario. The LCE
for both horizons shows the same behavior as the violation of GSLT
throughout the spacetime for $z$ and LCE parameter $\beta$ with
$h=3$. We have assumed $\beta>0$ \cite{e} which leads to positive
contribution to the entropy of the system. The values of these
correction parameters are not sensitive corresponding to the
obtained behavior of the rate of change of total entropy. However,
for very high value of $h$, the rate of change of total entropy
gives positive results for LCE while it becomes negative for PLCE.

Bamba et al. \cite{11++} studied the validity of GSLT in $f(T)$
gravity (with $F(T)=T+f(T)$) generally in thermal equilibrium for
apparent and event horizons. They assumed the basic requirement
$f_{TT}\ll1$ to hold the first law of thermodynamics in addition to
$f_T>0$ to constrain the PLCE and LCE parameters. They concluded
that for PLCE and LCE, GSLT satisfies for any value of correction
parameters for Hubble horizon. In case of event horizon, the
validity of GSLT depends upon the time derivative of event horizon
for both entropy corrections. These results hold regardless of any
choice of $f(T)$ model. However, we have obtained constraints on
correction parameters for a constructed $f(T)$ model to check the
validity of GSLT. Also, in a recent paper \cite{d}, we have
discussed the validity of this law incorporating the nonlinear
electrodynamics and dust matter in $f(T)$ gravity with two types of
scale factor for Hubble and event horizons. It was shown that this
law holds in this case only in the early universe and violates for
the present and future epochs for both horizons with power-law scale
factor. Here the PLCE provides the validity of GSLT for the same
scale factor and violation turns out for LCE. It is interesting to
mention here that all our results become equivalent to the results
of \cite{d} for zero entropy correction terms as well as magnetic
field.

\end{document}